\documentstyle[aps,preprint,epsf]{revtex}

\newcommand{\be}{\begin{equation}}
\newcommand{\ee}{\end{equation}}
\newcommand{\bea}{\begin{eqnarray}}
\newcommand{\beas}{\begin{eqnarray*}}
\newcommand{\eea}{\end{eqnarray}}
\newcommand{\eeas}{\end{eqnarray*}} 
\newcommand{\ba}{\begin{array}}
\newcommand{\ea}{\end{array}}
\newcommand{\bi}{\begin{itemize}}
\newcommand{\ei}{\end{itemize}}
\newcommand{\ben}{\begin{enumerate}}
\newcommand{\een}{\end{enumerate}}

\begin{document}


\title{Non SUSY and SUSY one--step Unification.}
\author{ Abdel P\'erez--Lorenzana$^{a}$
\and William A. Ponce$^{b}$ and Arnulfo Zepeda$^{a}$}
\address{
a Departamento de F\'{\i}sica,
Centro de Investigaci\'on y de Estudios Avanzados del I.P.N.\\
Apdo. Post. 14-740, 07000, M\'exico, D.F., M\'exico.\\
b Departamento de F\'\i sica, Universidad de Antioquia,  
 A.A. 1226, Medell\'\i n, Colombia.}
\date{\today}

\maketitle

\begin{abstract}
We explore the possibility of achieving one--step 
unification of the standard model 
coupling constants within non supersymmetric and supersymmetric  
gauge models, which at low energies have only
the standard particle content.
The  constraints are  the experimental values of 
$\alpha_{em}$, $\alpha_s$ and 
$\sin^2\theta_W$ at $10^2\, GeV$,  and  the lower bounds for FCNC and
proton decay rates. The analysis is done in a model independent
way.\\[1ex]
\end{abstract}
\vskip2em

It has been known for more than a decade\cite{sirlin} that if we let the
three gauge couplings $\alpha_i$ run through the ``desert" from
low to high energies, they do not merge together into a single point, as it
is shown in
Fig.1 ($\alpha_i=g_i^2/4\pi, i=1,2,3$ are the gauge couplings
for U(1)$_Y$, SU(2)$_L$, and SU(3)$_c$ respectively, the subgroups of the
standard model (SM) gauge group $G_{SM}=SU(3)_c\otimes SU(2)_L\otimes
U(1)_Y$). 

This result, which emerges from the accuracy measurements of the several
parameters in the SM done at the LEP machine at CERN\cite{lep}, claims
either for new physics at intermediate energy scales, or for new
approaches
to the unification problem. Conspicuously among the solutions calling for
new physics are those which introduce the minimal supersymmetric (SUSY)
partners of the SM fields at an energy scale of 1 TeV\cite{amaldi}.

The unification of the SM gauge couplings $\alpha_i,
\hspace{.2cm} i=1,2,3$ is properly achieved if the three values meet 
together  into a
common value $\alpha=g^2/4\pi$ at a certain energy scale $M>>10^2$ GeV,
where $g$ is the gauge coupling constant of the unifying group. Since
$G\supset G_{SM}$, the
normalization of the generators corresponding to the subgroups $U(1)_Y$,
$SU(2)_L$ and $SU(3)_c$ is in general different for each particular group
G, and therefore the SM coupling constants $\alpha_i$ differ at the
unification scale from $\alpha$ by numerical factors
$c_i \; (\alpha_i=c_i\alpha)$ which are pure rational numbers satisfying
$0<c_i\leq 1$\cite{few}. As a matter of fact, in Fig.1 the coupling
constants
have been normalized to the values $\{c_1,c_2,c_3\}=\{\frac{3}{5},1,1\}$,
which  are the normalization constants corresponding to the most popular
grand unified theories
(GUT) like SU(5), SO(10), E$_6$, etc.\cite{guts}.

But, can we normalize $\alpha_1$ to a different $c_1$ value in such a
way that $\alpha_1^{-1}(\mu)$ merge into the intersection of
$\alpha_2^{-1}(\mu)$ and $\alpha_3^{-1}(\mu)$ in the $\alpha_i^{-1}$-$\mu$
plane? 
The answer is yes. A value $c_1=39/50$ will
produce one-step unification of the three gauge couplings (see the dotted
line in Fig. 1), at an energy scale of the order of $10^{17}$ GeV, for the
standard containt of particles.
Unfortunately, the values $\{c_1,c_2,c_3\}=\{\frac{39}{50},1,1\}$ do not
correspond to any GUT known so far.

Our approach is to analyze in a  model independent way the solutions to
the renormalization group equations using $\{c_1,c_2,c_3\}$ as free
parameters, in order to look for non SUSY and SUSY GUT models able to
achieve one-step unification of the
SM gauge coupling constants, consistent with the low energy
phenomenology. As we already know, the value of the SM
coupling constants at the $m_Z$ scale and the bounds on the proton
life time, rule out models like 
minimal $SU(5)$, and other models that contain minimal $SU(5)$ as an
intermediate stage in their symmetry braking chain (sbc). 
To simplify matters we use for $c_3$ only the values 1 and $\frac{1}{2}$.  
($c_3=1$ for models which contain $SU(3)_c$ embedded into a simple
subgroup of G, and $c_3=\frac{1}{2}$ for models which contain $SU(3)_c$
embedded into the chiral color extension $SU(3)_{cL}\otimes
SU(3)_{cR}\subset G$\cite{2su3}).


In a field theory, the coupling constants are defined as effective values
including loop corrections of the particle propagators according to
the renormalization group equations. They are therefore energy scale
dependent. In the  modified minimal substration scheme
($\overline{MS}$)~\cite{ms}, which we adopt in what follows, the
one--loop renormalization group equations (rge) are
\be
\mu{d\alpha_i\over d\mu} \simeq -b_i \alpha_i^2, \label{rge}
\ee
where $\mu$ is the energy at which the coupling constants $\alpha_i$ 
are evaluated. The constants $b_i$ are completely determinated by the
particle content in the model by
\be
4\pi b_i = {11\over 3} C_i(vectors) - {2\over 3}C_i(fermions)
-{1\over 3}C_i(scalars), \label{bethas}
\ee
being $C_i(\cdots)$ the index of the representation to which the $(\cdots)$
particles are assigned, and where we are considering Weyl fermion and
complex scalar fields~\cite{bs}. The boundary conditions at the
scale $m_Z\simeq 10^2 GeV$
 for these equations are determined by the relationships
\be
\alpha^{-1}_{em} = \alpha_1^{-1}+ \alpha_2^{-1},  \quad\mbox{and}\quad
\tan^2\theta_W = {\alpha_1\over\alpha_2}, \label{rel1}
\ee
where $\alpha_{em}=e^2/4\pi$ ($e$ the electric charge), and by the experimental
values 
\bea
\alpha^{-1}_{em}(m_Z) &=& 127.90 \pm 0.09~\cite{lep,pdg,aem},\nonumber\\  
\sin^2\theta_W (m_Z)&=& 0.2312 \pm 0.00017~\cite{lep,pdg} \quad
\label{exp}\mbox{and}\\
\alpha_3(m_Z) &=& \alpha_s = 0.1191\pm 0.0018~\cite{pdg}.\nonumber
\eea
which are  the updated world average of all current data.

From eq.(\ref{rel1}), which are valid at all energy scales, it follows that at 
the unification scale $M$, the value $\sin^2\theta_W$ is 
\be
\sin^2\theta_W(M) = {\alpha_{em}(M)\over\alpha_2(M)} = {c_1\over c_1 +
c_2}.
\ee

For the non SUSY case, under the assumptions that only the three standard 
families of particles are light, and using the decoupling theorem~\cite{appel}, 
the solution to the rge can be written as
\be
\alpha_i^{-1}(m_Z) = {1\over c_i}\alpha^{-1}- b_i(F,H)\, \ln\left({M\over
m_Z}\right), \label{system}
\ee
where
\be \label{bi}
2 \pi \left( \ba{c} b_1\\[1ex] b_2\\[1ex] b_3 \ea \right) = 
 \left( \ba{c} 0\\[1ex] 22\over 3 \\[1ex] 11 \ea \right) - 
 \left( \ba{c} 20\over 9 \\[1ex] 4\over 3 \\[1ex] 4 \over 3 \ea \right) F - 
 \left( \ba{c}  1\over 6 \\[1ex]   1\over 6 \\[1ex] 0 \ea \right) H.
\ee
$F=3$ is the number of light families and $H=1$ is the number of low energy 
complex Higgs doublets. 
(Notice that we are not including in Eq.(\ref{bi}) the
normalization factor $3\over 5$ into $b_1$ coming from
the $SU(5)$ theory, and wrongly included in some general discussions.) 
Once the set $\{c_1,c_2,c_3\}$ is provided for a particular group $G$, the
former equations constitute a system of three equations with two unknowns:
$\alpha$ and $M$ ($\alpha_i(m_Z), i=1,2,3$ are obtained from the values
presented in (\ref{exp})). So, a consistent check of the GUT hypothesis is
in principle possible.

Our approach now is the following\cite{abdel}: we consider the system of three
equations (\ref{system}) with the three unknowns $\alpha, M$ and $H$, each one of
the unknowns a function of the parameters $\{c_1,c_2,c_3\}$; we solve for the
three unknowns as functions of $c_i,i=1,2,3$ and draw curves for physical values
of $M$ in a 3 dimensional cartesian space, where the coordinate axis are
provided by the set $\{c_1,c_2,c_3\}$, (actually, since $c_3=1,\frac{1}{2}$ we
considered only two dimensional spaces with axis $\{c_1,c_2\}$, projected into the
planes $c_3=1$ and $c_3=\frac{1}{2}$). 

The physical values of $M$ are provided by experimental and theoretical
bounds in the following way: first, the unification scale $M$ must be
lower than the Plank scale $M_{P}\sim G_N^{-1/2}\sim 10^{19}GeV$; second,
it must be greater than $10^5GeV$ in order to cope with experimental
bounds on FCNC~\cite{pdg}. Finally, since some models predict proton
decay, and the experimental bound for the proton life time $\tau_p$ is
$\tau_{p\rightarrow e\pi}\sim M^4 >10^{33}$ Yrs~\cite{kamioca}, then $M$
must be greater than $10^{16}$ GeV if the proton is unstable in the model
under consideration. Hence, in the analysis we have to consider two
different zones in the $c_1-c_2$ plane, given by $10^{16}$GeV $<M<M_{P}$
and $10^5$GeV $\leq M \leq 10^{16}$GeV, which admit and does not admit
proton decay respectively. Also, since $b_3>0$ and $b_1<0$ always,
$\alpha_1(m_Z)<\alpha<\alpha_s(m_Z)/c_3$ then $\alpha$, $\ln (M/m_Z)$ and
$H$ should be finite, and there is an upper bound $H_{max}$ which
represents the maximum number of low energy Higgs doublets allowed.
Therefore, $0\leq H\leq H_{max}$. These bounds limit the region in the
$\{c_1,c_2\}$ plane where the coupling constant one-step unification is
possible and consistent with the experimental data and theoretical
constraints.

The solutions to Eqs. (\ref{system}) for $\alpha, H$ and $M$ as functions of
$c_i$ are:  
\be
\alpha^{-1} =c_1c_2c_3\cdot 
{(\alpha_1^{-1}-\alpha_2^{-1})(99 - 12F) + \alpha_3^{-1}(8F + 66)
\over c_1c_2(8F + 66) + c_1 (c_1-c_2)(12F - 99)},
\ee
\be
H = {2\over 3}\cdot {c_2(\alpha_1^{-1}c_1 - \alpha_3^{-1}c_3)(66-12F) + 
c_3(\alpha_1^{-1} c_1 - \alpha_2^{-1}c_2)(12F-99) +
20c_1(\alpha_2^{-1}c_2 - \alpha_3^{-1}c_3)
\over c_1c_2(\alpha_1^{-1}-\alpha_2^{-1}) + \alpha_3^{-1}c_3 (c_1-c_2)},
\ee
\be
\ln\left({M\over m_Z}\right) = 
18\pi\cdot {c_1c_2(\alpha_1^{-1}-\alpha_2^{-1}) + \alpha_3^{-1}
c_3 (c_1-c_2)\over c_1c_2(8F + 66) + c_1 (c_1-c_2)(12F - 99)}.
\ee
From these expressions, the limited region obtained for values of $c_1$ and
$c_2$ that give unification is plotted in figure 2 for $c_3=1$ and in 
figure 3 for $c_3={1\over2}$, where we used $F=3$ for three light families,
and central values for  $\alpha_s(m_Z)$, $\alpha_{em}(m_Z)$ and
$\sin^2\theta_W(m_Z)$. Let us analyze those graphs:

\noindent
{\bf Analysis of Fig. 2}: It corresponds to GUT groups with vector-like  
color symmetries. The allowed region of parameters
$\{c_1,c_2\}$ lies inside the lines $M=10^5$ GeV, $H=0$, and $c_2=1$.
There is a maximum unification mass scale given by 
$M\leq 10^{17.5}$ GeV $< M_P$ and the number of Higgs field doublets
allowed is such that $0<H\leq 91$ in general, but if the proton does decay
in the context of the GUT model then $0<H\leq 2$.
The implications are:\\
1: For SU(5)\cite{georgi}, SO(10) \cite{so10}, E$_6$ \cite{e6}, and SO(18)
\cite{so18}, 
$\{c_1,c_2\}=\{\frac{3}{5},1\}$ and proton decay is always present.
The point lies inside the allowed zone,
but in a region where $M\simeq 10^{13}$GeV in conflict with the bounds for
proton decay. Since $SU(5)$ allows only the one step sbc 
$SU(5)\stackrel{M}{\longrightarrow}SM$, $SU(5)$ is ruled out in
general (not
only minimal $SU(5)$ but also all the possible extensions which include
arbitrary representations of Higgs field multiplets). 
The one-step sbc for SO(10), E$_6$ and SO(18) are also ruled out by the same 
reason.\\
2: For SU(4)$_c\otimes SU(2)_L\otimes SU(2)_R$\cite{patis}, and 
$[SU(3)]^3\times Z_3$\cite{rug}, 
$\{c_1,c_2\}=\{\frac{3}{5},1\}$ again. In those models the 
proton can not decay via leptoquark gauge bosons (see the first paper in
~\cite{patis} and the last paper in \cite{rug}), but it can decay via Higgs 
field scalars. So, the one
stage breaking of those models is not ruled out as long as one can break the
symmetry using scalars which do not break spontaneously the baryon
quantum number. The GUT scale for those models is $M\simeq 10^{13}$ GeV
and 
the number of Higgs field doublets is $H=7\pm 1$.\\
3: For $[SU(6)]^3\times Z_3$ \cite{3su6}, $\{c_1,c_2\}=\{{3\over 14},
{1\over 3}\}$ which lies outside the allowed zone.
The one stage sbc is ruled out for this
model (the two stage sbc is presented in some papers of
Ref.~\cite{3su6}).\\
4: For E$_7$ \cite{e7},
$\{c_1,c_2\}=\{\frac{3}{2},\frac{1}{2}\}$\cite{few}, values way out the
allowed region. The one stage sbc is also ruled out for this model.

\noindent
 {\bf Analysis of Fig. 3}: It corresponds to GUT groups with chiral color
symmetries. The allowed region of parameters 
$\{c_1,c_2\}$ lies inside the lines $M=10^5$ GeV, $H=0$, $c_2=1$, and 
$M=M_P=10^{19}$ GeV. There is not maximum  bound for a unification
mass scale, 
and the allowed number of Higgs field doublets is $0<H\leq 136$
in general, but if the proton does decay in the context of the GUT model
then $0<H\leq 28$. The implications for some specific models
are:\\
1: For $SU(5)\otimes SU(5)$ \cite{2su5}, $\{c_1,c_2\}=\{{3\over 13},1\}$
which lies inside the allowed zone but in a region where $M<<10^{16}$
GeV, in serious conflict with bounds for proton decay. The one step
sbc for this model is ruled out.\\
2: For $[SU(6)]^4\times Z_4$ \cite{4su6}, $\{c_1,c_2\}=\{{3\over 19} ,
{1\over 3}\}$ which lies inside the allowed zone (the proton is stable 
in the context of this model). So, the one stage sbc for this model is
allowed (it is presented in Ref. \cite{4su6}), the unification scale is
$M\sim 10^7$ GeV, and the number of low energy Higgs field doublets is
$H=2\pm 1$.

\noindent
{\bf NOTE} Comparing Fig. 2 with Fig. 3 we conclude that one-step family
unification is more likely achieved if $SU(3)_c$ is embedded into the 
chiral color group $SU(3)_{cL}\otimes SU(3)_{cR}$.  


If SUSY plays a role in our low energy world, its most likely mass scale
is consider to  be $M_S\sim 1$ TeV \cite{susy}
(the inclusion of the low energy SUSY threshold correction and others,
could 
be important, in such a way that the effective mass scale $M_S$ should be
taken lower than 1 TeV, may be as low as $m_Z$~\cite{polonsky}).
In what follows we assume that
below $M_S$ we
have only the SM physics and above $M_S$ the minimal supersymmetric
standard model (MSSM) manifest itself, up to the unification scale $M$. 
In this context, the solution to the rge with the mass hierarchy
$m_Z<M_S<M$ is given by
\be
\alpha_i^{-1}(m_Z)=\alpha_i^{-1}(M)-b_i(F,H){\rm ln}({M_S\over m_Z}) -
b_i^{SS}(F,H){\rm ln}({M\over M_S})
\ee
where $b_i^{SS}$ are the contributions to the beta function of the MSSM, 
and $b_i$ given by Eq. (\ref{bi}), the contributions of the SM.
Again $b_i$ and 
 $b_i^{SS}$ depend on the number of low energy families $F$ and
Higgs field doublets $H$, decoupling in each case the extra massive
particles according to the decoupling theorem~\cite{appel}. The analysis
now produces
\be \label{biss}
2 \pi \left( \ba{c} b_1^{SS}\\[1ex] b_2^{SS}\\[1ex] b_3^{SS} \ea \right) = 
 \left( \ba{c} 0\\[1ex] 6 \\[1ex] 9 \ea \right) - 
 \left( \ba{c} 10\over 3 \\[1ex] 2 \\[1ex] 2 \ea \right) F
- 
 \left( \ba{c}  1\over 2 \\[1ex]   1\over 2 \\[1ex] 0 \ea \right) H,
\ee
where again the $3\over 5$ normalization factor comming from SUSY
$SU(5)$ has not been included in $b_1^{SS}$.

Repeating the analysis done for the non SUSY case we get
the results plotted in Figs. 4 and 5. From the graphs we see that the
existence of SUSY partners not only changes the shape of the allowed
regions but has deeper consequences, as can be seen from Fig. 4 where the
point $\{c_1,c_2\}=\{{3\over 5}, 1\}$ associated with $SU(5)$ and
related models now fits into the allowed region, a well known 
result from a related analysis\cite{amaldi}.
For the several models we have studied, our conclusions for the SUSY case
(from Figs. 4 and 5) are:\\
1- SUSY models with one step sbc allowed: SU(5), SO(10), E(6), 
SU(4)$\otimes$SU(2)$_L\otimes$SU(2)$_R$,
$[SU(3)]^3\times Z_3$, and [SU(6)]$^4\times Z_4$.\\
2-SUSY models with one step sbc ruled out:
[SU(6)]$^3\times Z_3$, E$_7$, and SU(5)$\otimes$SU(5).


Our conclusion is that it is always possible for a certain class of GUT
models  to
achieve a one--step unification, both in the non SUSY and in the SUSY
cases. For some SUSY models this result was known before\cite{amaldi}, 
but for the non SUSY cases the result is new and not trivial.

Our analysis has been done including only the one--loop beta function. If
one uses a two--loop beta function, then one-loop threshold corrections
(which are model dependent) must be included, and the experimental errors
of the SM group coupling constants (specially for $\alpha_s$) must be
taken into account since all of them are of the same order of magnitud
(typically of the order of 2 to 8 \%). The inclusion of those
contributions to the gauge coupling constants do not change our
conclusions. (In Ref. \cite{polonsky} such analysis is presented for the
SM, the MSSM and GUT SU(5)).

Finally notice that the gauge coupling constants $g_i$ of the three SM
interactions are related at the GUT scale by the relationship $c_1^{-1}
g^2_1= c_2^{-1} g^2_2= c_3^{-1} g^2_3$, a result which resemble the string
gauge coupling unification. Indeed, defining $c_i^{-1} = \kappa_i$, the
affine level (or Kac-Moody level)  at which the group factor $G_i$ is
realized in the effective four dimensional string, we get the string
coupling relation~\cite{gins} $\kappa_1 g_1^2=\kappa_2 g^2_2=\kappa_3
g^2_3$. This analogy, together with the results presented here, should
provide further insight into the String-GUT problem\cite{iban}, for the
SUSY and the non SUSY string unification. As a matter of fact, Fig. 1
above suggests a string unification without supersymmetry for the
Kac-Moody levels $\kappa_1 = \kappa_Y = 1.28$ and $\kappa_2= \kappa_3 = 1
$ (see also Refs. \cite{dien}). 

\noindent
{\bf Acknowledgements}.  
We thank F. del Aguila for sharing with us his wisdom, G. Aldazabal
for several comments on String-GUT, and K. Dienes for a written
communication.  
This work was partially supported by CONACyT, M\'exico and COLCIENCIAS,
Colombia.


\newpage

\figure{
\centerline{
\epsfxsize=300pt
\epsfbox{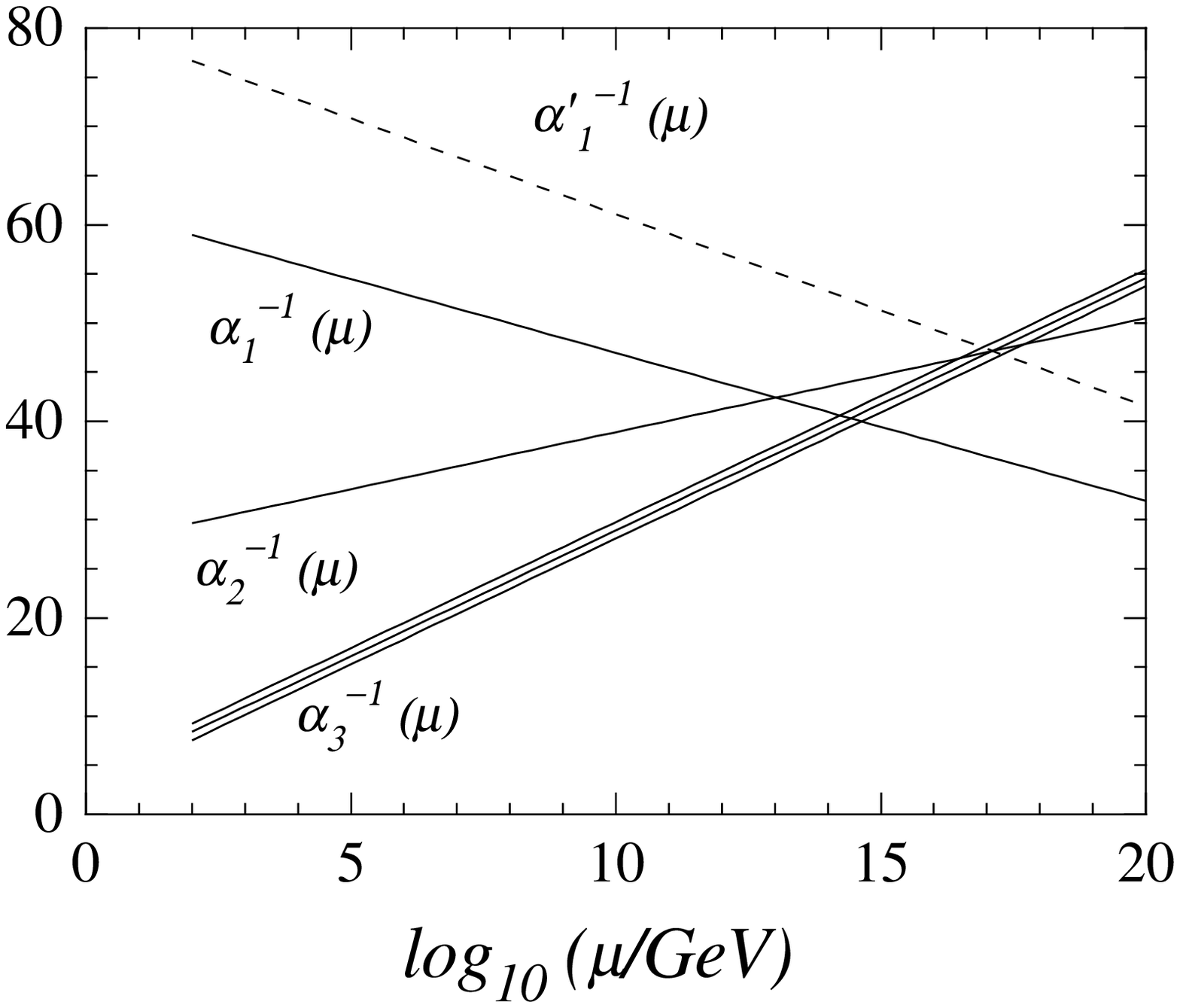}
}
\caption{One-loop evolution of the gauge couplings with the (non-susy) 
Standard Model. Here $\alpha_1\equiv (5/3) \alpha_Y$ for the solid line
and $\alpha'_1\equiv (50/39) \alpha_Y$ for the dotted line, where
$\alpha_Y$ is the hypercharge coupling.}
\label{run}
 }
\vskip1em

\figure{
\centerline{
\epsfxsize=300pt
\epsfbox{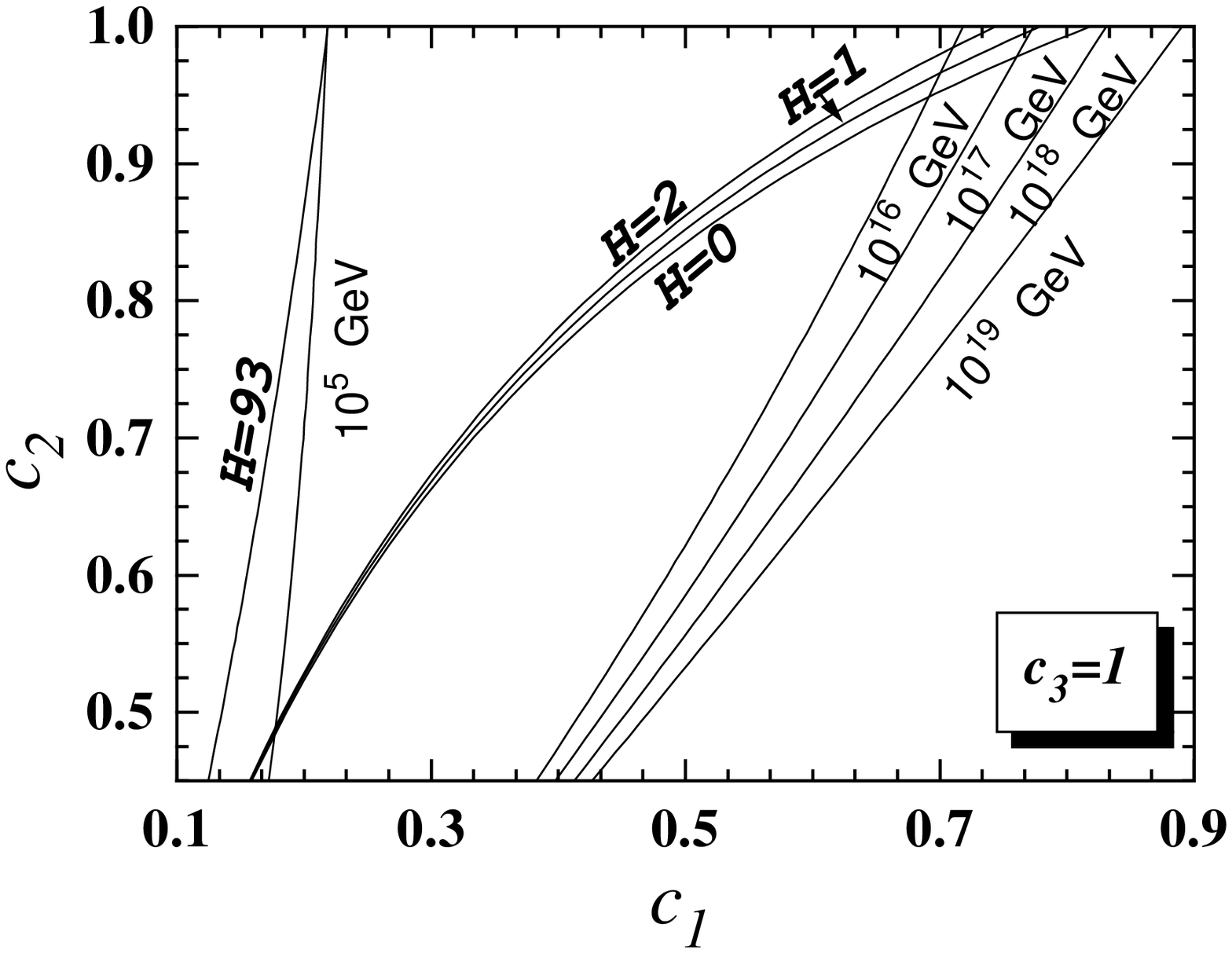}
}
\caption{Plots for some values of H and M for the non chiral color models. The
bounds in $c_1$ and $c_2$, impose at once for $\alpha$ the bounds $16.0921<
\alpha^{-1} < 48.0186$.}
\label{fig1}
}
\vskip1em
\figure{
\centerline{
\epsfxsize=300pt
\epsfbox{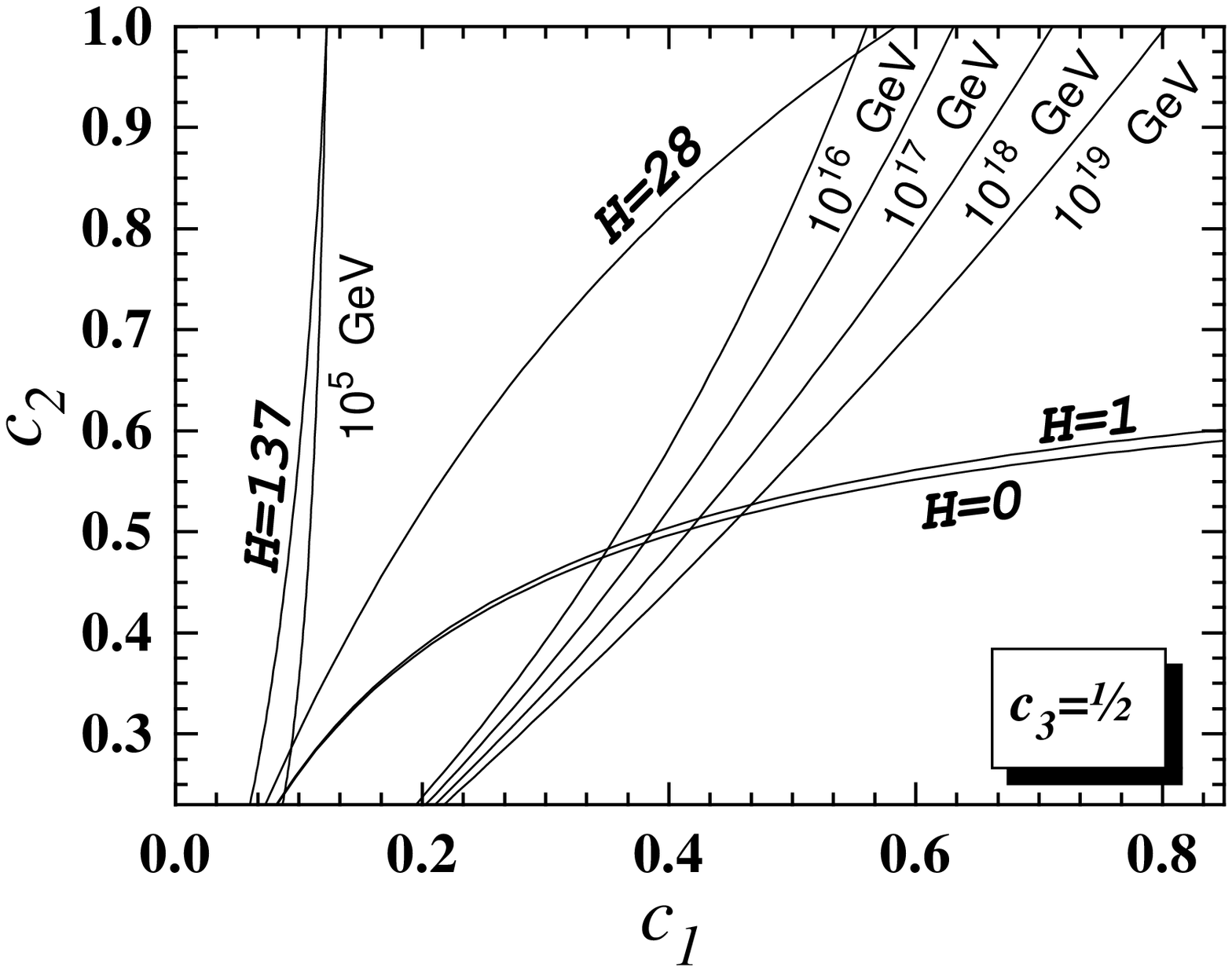}
}
\caption {Plots for some values of $H$ and $M$ for GUT containing the chiral
color extension. In this case we have $8.0461< \alpha^{-1}< 26.003$.}
\label{fig2}
}
\vskip 1em

\figure{
\centerline{
\epsfxsize=300pt
\epsfbox{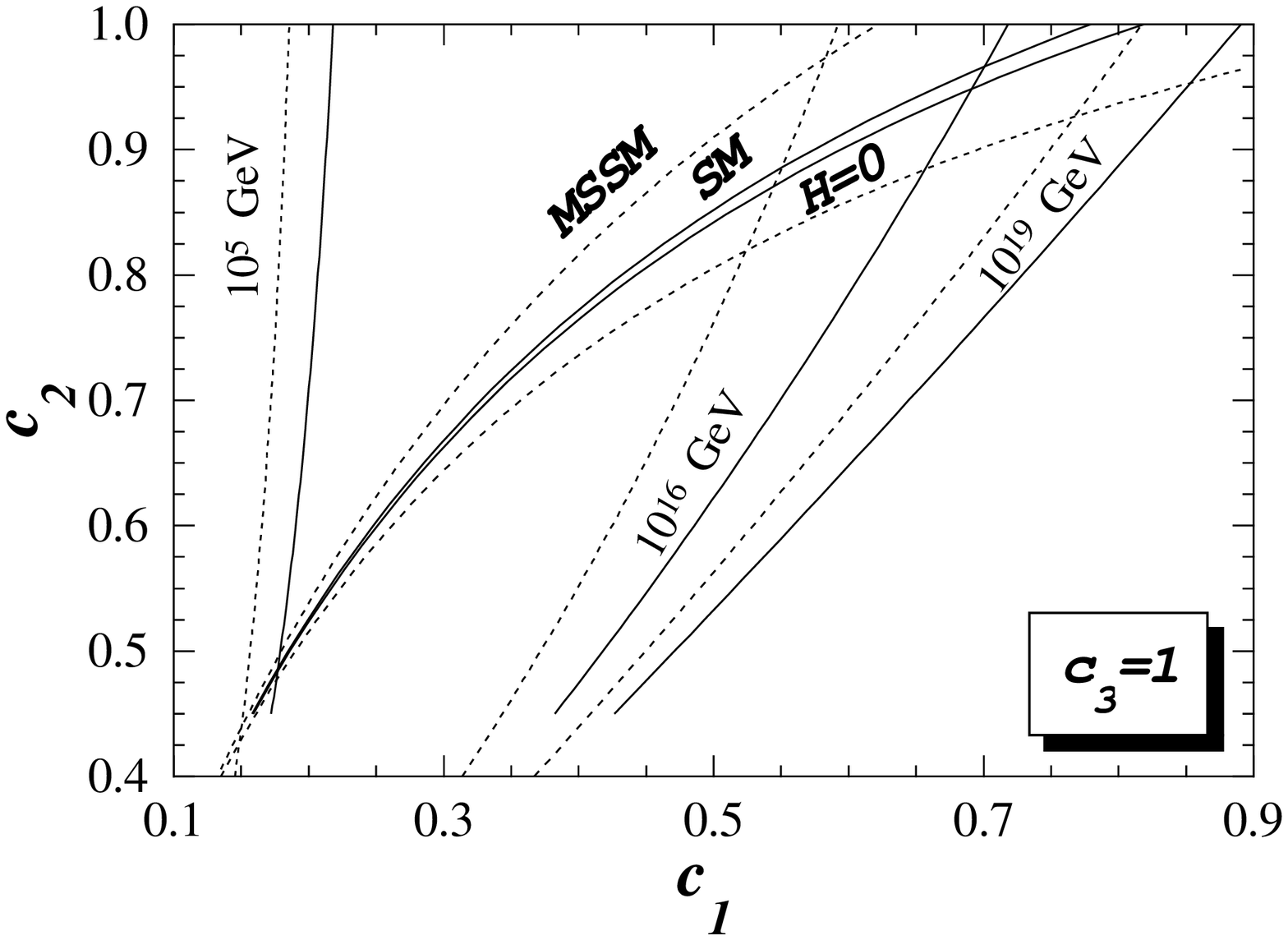}
}
\caption{Comparing between the fits for H and M in SUSY and  non SUSY   
unification, for the non chiral color models. The dashed lines stand for the
SUSY case. Now, $13.16\leq \alpha^{-1} \leq 28.552$.}
\label{fig3}
}
\vskip1em
\figure{
\centerline{
\epsfxsize=300pt
\epsfbox{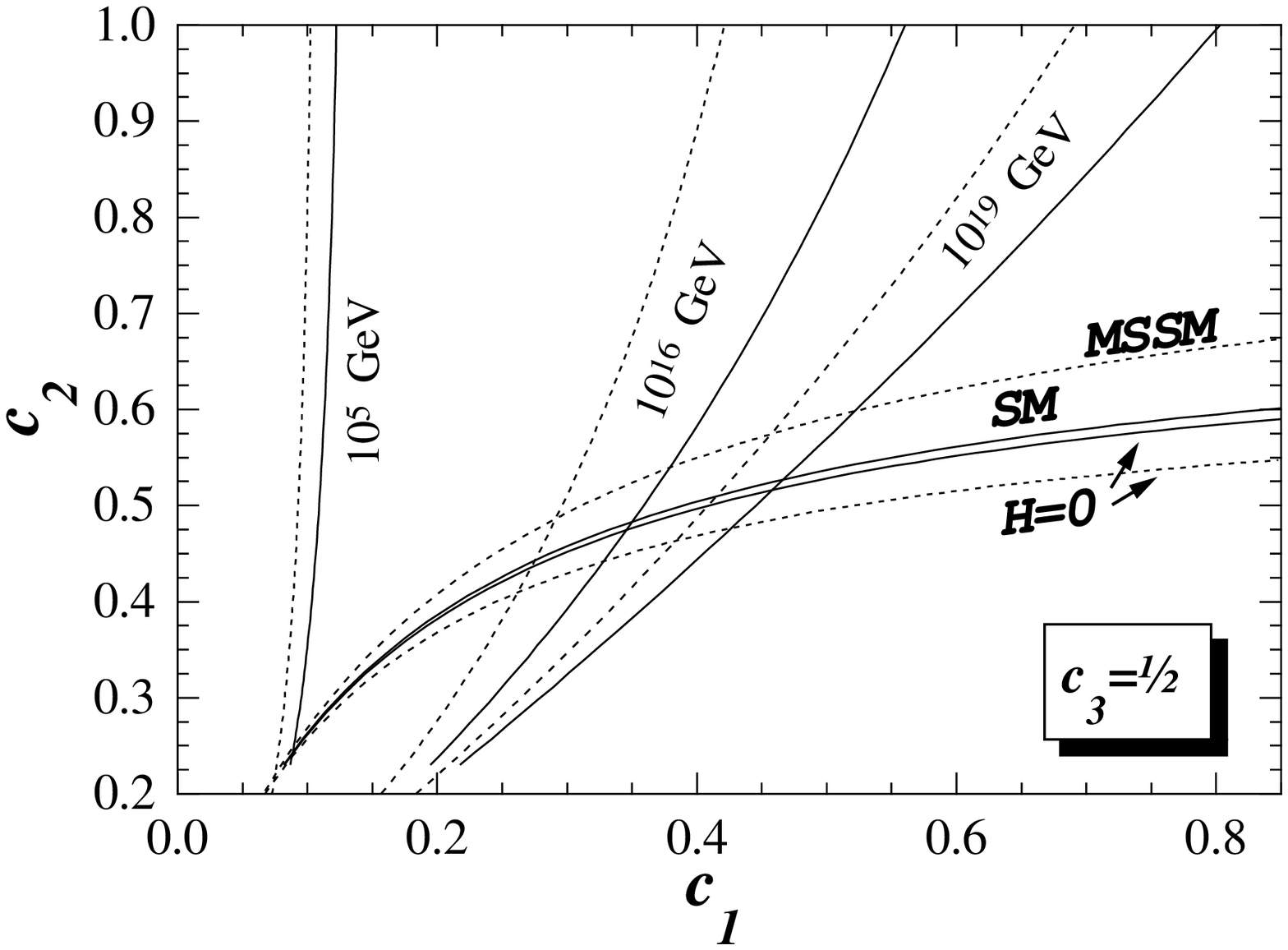}   
}

\caption{Comparing between the bounded surface for SUSY and non SUSY simple
unification, containing the chiral color extension. Again, the dashed lines
correspond to the SUSY fits. In this case $6.58\leq \alpha^{-1}\leq
14.276$.}
\label{fig4}
}
\vskip 1ex


\begin{thebibliography}{99}
\bibitem{few}
For all the models $c_2,c_3\leq 1$, but we have found a few models with
$c_1>1$ as E$_7$ for example\cite{e7}.
\end{thebibliography}
\end{document}